\documentclass[
reprint,
superscriptaddress,
amsmath,
amssymb,
aps,
prl,
nofootinbib,
]{revtex4-1}

\usepackage{graphicx}
\usepackage{dcolumn}
\usepackage{bm}
\usepackage{physics}
\usepackage{siunitx}
\usepackage[colorlinks=true, linkcolor=blue, citecolor=blue, urlcolor=blue]{hyperref}
\usepackage{makecell} 
\usepackage{booktabs} 
\usepackage{mathtools} 
\DeclareSIUnit\angstrom{\text{Å}}
\usepackage{array} 
\usepackage{braket}
\usepackage{acro} 
    \DeclareAcronym{PL}{
      short = PL,
      long  = photoluminescence
    }
    \DeclareAcronym{PLE}{
      short = PLE,
      long  = photoluminescence excitation
    }
    \DeclareAcronym{ED}{
      short = ED,
      long  = electric dipole
    }
    \DeclareAcronym{MD}{
      short = MD,
      long  = magnetic dipole
    }
    \DeclareAcronym{FWHM}{
      short = FWHM,
      long  = full width at half maximum
    }

\begin{document}

\title{Spectroscopic Investigations of Multiple Environments in Er:CaWO\texorpdfstring{$_4$}, through Charge Imbalance}

\author{Fabian Becker}
    \affiliation{TUM School of Computation, Information and Technology, Technical University of Munich, 80333 Munich, Germany}
    \affiliation{Walter Schottky Institute, Technical University of Munich, 85748 Garching, Germany}
    \affiliation{Munich Center for Quantum Science and Technology (MCQST), 80799 Munich, Germany}
\author{Catherine L. Curtin}
    \affiliation{TUM School of Computation, Information and Technology, Technical University of Munich, 80333 Munich, Germany}
    \affiliation{Walter Schottky Institute, Technical University of Munich, 85748 Garching, Germany}
    \affiliation{Munich Center for Quantum Science and Technology (MCQST), 80799 Munich, Germany}
\author{Sudip KC}
    \affiliation{TUM School of Computation, Information and Technology, Technical University of Munich, 80333 Munich, Germany}
    \affiliation{Walter Schottky Institute, Technical University of Munich, 85748 Garching, Germany}
    \affiliation{Munich Center for Quantum Science and Technology (MCQST), 80799 Munich, Germany}
\author{Tim Schneider}
    \affiliation{TUM School of Computation, Information and Technology, Technical University of Munich, 80333 Munich, Germany}
    \affiliation{Walter Schottky Institute, Technical University of Munich, 85748 Garching, Germany}
    \affiliation{Munich Center for Quantum Science and Technology (MCQST), 80799 Munich, Germany}
\author{Lorenz J. J. Sauerzopf}
    \affiliation{TUM School of Computation, Information and Technology, Technical University of Munich, 80333 Munich, Germany}
    \affiliation{Walter Schottky Institute, Technical University of Munich, 85748 Garching, Germany}
    \affiliation{Munich Center for Quantum Science and Technology (MCQST), 80799 Munich, Germany}
\author{Ibrahim Elzeiny}
    \affiliation{TUM School of Computation, Information and Technology, Technical University of Munich, 80333 Munich, Germany}
    \affiliation{Walter Schottky Institute, Technical University of Munich, 85748 Garching, Germany}
    \affiliation{Munich Center for Quantum Science and Technology (MCQST), 80799 Munich, Germany}
\author{Kai M\"uller}
    \affiliation{TUM School of Computation, Information and Technology, Technical University of Munich, 80333 Munich, Germany}
    \affiliation{Walter Schottky Institute, Technical University of Munich, 85748 Garching, Germany}
    \affiliation{Munich Center for Quantum Science and Technology (MCQST), 80799 Munich, Germany}

\date{\today}

\begin{abstract}
We present a detailed spectroscopic study of the $^4\mathrm{I}_{13/2}$ and $^4\mathrm{I}_{15/2}$ Er$^{3+}$ multiplets of Er:CaWO$_4$ grown without a co-dopant. Using photoluminescence and photoluminescence excitation measurements, we find multiple environments into which the Erbium ions are incorporated in the crystal. For the four most prevalent environments, we provide a complete assignment of the $^4\mathrm{I}_{13/2}$ and $^4\mathrm{I}_{15/2}$ sublevel energies. Polarization-dependent absorption and emission spectroscopy allow us to identify the irreducible representations of the states and determine the dipole nature of transitions. In addition, we compare the different relaxation paths within $^4\mathrm{I}_{13/2}$ and $^4\mathrm{I}_{15/2}$ multiplets via line broadening and lifetime measurements. Finally, spectral, lifetime, and polarization measurements indicate similarities between the studied environments.
\end{abstract}
\maketitle
\section{Introduction}
Quantum networks are a cornerstone technology for advancing quantum communication and distributed quantum computing \cite{Awschalom.2021, Yu.2020}. Central to the success of these networks are quantum repeaters, which overcome the exponential decay of entanglement transmission over long distances using quantum memories \cite{Briegel.1998, Bussieres.2013, Lvovsky.2009}, and microwave to optical transducers transferring microwave qubits to the optical regime \cite{Yu.2020, Xie.2021}. Recent studies highlighted the pivotal role of atom-like defect centers in solid-state platforms with long coherence times for such quantum applications \cite{Awschalom.2018, Kanai.2022}. Ferrenti et al. (2020) \cite{Ferrenti.2020} summarized the requirements for solid-state hosts of quantum defects to be intrinsically diamagnetic, have a large band gap, be free of paramagnetic impurities or unwanted defects, existence of a known method of epitaxial growth, absence of host nuclei with nonzero magnetic moments and finally no crystalized phases in polar space groups. Under these considerations, the crystal CaWO$_4$ was identified as one of the most suitable candidates for hosting quantum defects \cite{Kanai.2022, Ferrenti.2020}. 
In particular, the potential of the rare-earth ion Erbium (Er$^{3+}$) to maintain long spin coherence times in low nuclear spin environments while maintaining its uniquely suited emission into the low loss window of optical fibers at $\SI{1.5}{\mathrm{\mu} m}$ \cite{Reiserer.2022} underscores its suitability for quantum applications. These properties are essential for minimizing decoherence and ensuring reliable quantum state transfer \cite{LagoRivera.2021, Rancic.2018}. Er:CaWO$_4$ attracted increasing attention due to its beneficial spin material system~\cite{LeDantec.2021, Ourari.192023, Uysal.6102024, Rancic.2022, Wang.2023} leading to a measured spin coherence time of $\SI{23}{ms}$ \cite{LeDantec.2021}. Moreover, key milestones for quantum applications such as indistinguishable single photon generation~\cite{Ourari.192023} and spin-photon entanglement~\cite{Uysal.6102024} were lately demonstrated. In this matrix Er$^{3+}$ is replacing Ca$^{2+}$, leading to a charge imbalance which causes additional transitions to appear \cite{Cornacchia.2007, Nassau.1963, Enrique.1971}. In previous strategies for laser applications, alkali metals were co-doped to minimize additional lines. This co-doping adds another source of spin noise due to the non-zero nuclear spin and potentially reduces spin coherence. To further understand and use this charge-imbalanced system, detailed spectroscopic studies are essential yet up to now missing.

In this paper, we investigate the charge imbalanced Er:CaWO$_4$ and provide the full $^4\mathrm{I}_{13/2}$ and $^4\mathrm{I}_{15/2}$ energy structure of the four most visible environments. Furthermore, we present the orientations of their absorption and emission dipoles and, from these, assign their irreducible representations and transition selection rules assuming an S$_4$ symmetry. Additionally, we highlight differences in relaxation and depopulation paths.
\section{CaWO\texorpdfstring{$_4$}, Host} \label{Sec_Host}
\begin{figure}[t]
    \centering
    \includegraphics[width=\dimexpr0.89\columnwidth\relax,trim=20 45 500 20,clip]{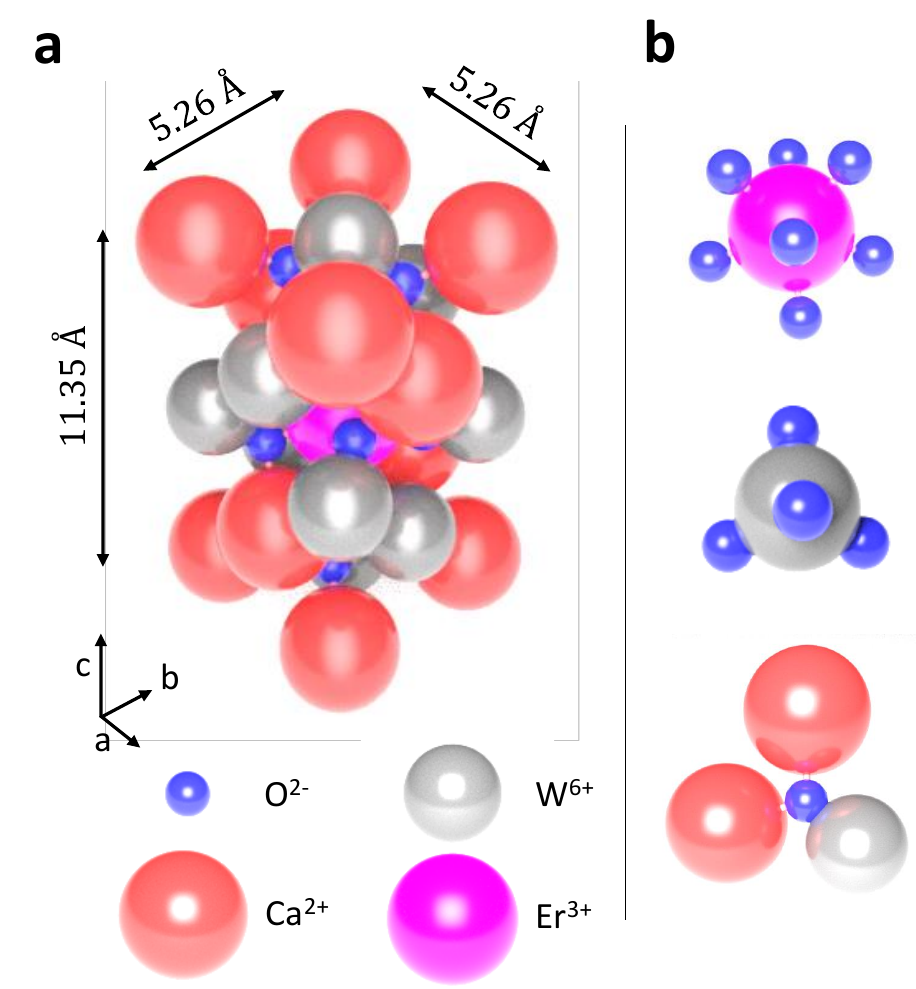}
    \caption{Crystalline CaWO$_4$ host structure with incorporated Er$^{3+}$. \textbf{a}: Schematic representation of the tetragonal I$4_1$/a group. The visualization of covalent atomic radii and distances is to scale. (data taken from \cite{MatProj_CaWO4} and \cite{Haynes.op.2014}); \textbf{b}: Crystal geometries surrounding the different species.}
    \label{Fig_Crystal}
\end{figure}

The uni-axial CaWO$_4$ matrix consists of three species. Based on the natural abundance of each element, the only potentially isotopes present with non-zero magnetic moments are $^{43}$Ca ($\SI{0.135}{\%}$) and W$^{183}$ ($\SI{14.31}{\%}$) \cite{Haynes.op.2014}. Figure~\ref{Fig_Crystal}a shows the tetragonal crystal structure of CaWO$_4$ with the incorporated Er$^{3+}$ on a Ca$^{2+}$ site with S$_4$\footnote{We use S$_4$ to denote the point group and $S_4$ for the symmetry element.} point group symmetry \cite{Mims.1965, Kiel.1970}. The size of the unit cell is based on the literature values given in \cite{MatProj_CaWO4}. To visualize the size of the individual atoms, the covalent atomic radius is used.\footnote{Covalent radii according to \cite{Haynes.op.2014}: $r_{\text{O}}=\SI{0.64}{\angstrom}$; $r_{\text{Ca}}=\SI{1.74}{\angstrom}$; $r_{\text{Er}}=\SI{1.77}{\angstrom}$; $r_{\text{W}}=\SI{1.50}{\angstrom}$. Ionic radii according to \cite{Haynes.op.2014}: $r_{\text{O}^{2-},CN=2}=\SI{1.21}{\angstrom}$; $r_{\text{Ca}^{2+},CN=8}=\SI{1.12}{\angstrom}$; $r_{\text{Er}^{3+},CN=8}=\SI{1.00}{\angstrom}$; $r_{\text{W}^{6+},CN=4}=\SI{0.42}{\angstrom}$} Figure~\ref{Fig_Crystal}b highlights the crystalline environment of each species. Ca$^{2+}$ or the incorporated Er$^{3+}$ is bound in a deltahedron to eight O$^{2-}$ ions with four bond lengths of $\SI{2.44}{\angstrom}$ and four of $\SI{2.47}{\angstrom}$. Tungsten is bound in a tetrahedron to its four surrounding oxygen ions with bond length of $\SI{1.80}{\angstrom}$. Oxygen is bound in a distorted geometry to two equally distant Ca$^{2+}$ and one Tungsten ion. Moreover, CaWO$_4$ is a crystal with $c=\SI{11.35}{\angstrom}$ and $a=b=\SI{5.26}{\angstrom}$, a refractive index of $\mathrm{n(o)=}\SI{1.88}{}$ and $\mathrm{n(e)=}\SI{1.90}{}$ at $\SI{1550}{nm}$~\cite{Polyanskiy.2024} and a band gap of $\SI{4.34}{eV}$ \cite{MatProj_CaWO4}.
Er$^{3+}$ is known to replace Ca$^{2+}$ in two paramagnetic sites. Both differ by inversion of the environment through the S4 symmetry site and are energetically indistinguishable without an applied electric field \cite{Mims.1965}. In addition, Kiel et al. (1970) \cite{Kiel.1970} found good agreement for the g tensor symmetry for $\mathrm{Yb^{3+}}$ between experiment and calculation by assuming an incorporation at the normal $\mathrm{Ca^{2+}}$ site without nearby charge compensation. Furthermore, Nassau, et al. (1963) \cite{Nassau.1963} studied the incorporation of rare-earth ions and charge compensation in CaWO$_4$ using Nd$^{3+}$ and Ce$^{3+}$. They related additional peaks to vacancy formation, that minimizes when Na$^{+}$ was co-doped. Additional interstitial incorporation of the Lanthanoids seemed unlikely, as CaWO$_4$ (scheelite) is very compact. Replacement of a W$^{6+}$-ion for charge compensation was expected to be insignificant. However, co-dopants of the alkali metal family (monovalent ions) would add magnetic noise due to their non-zero nuclear spin and potentially reduce spin coherence. Thus, they are not favorable for quantum platforms \cite{Ferrenti.2020}. 
\section{Site-Selective Photoluminescence} \label{Sec_PLE}
\begin{figure}[t]
    \centering
    \includegraphics[width=\columnwidth,trim=520 45 0 40,clip]{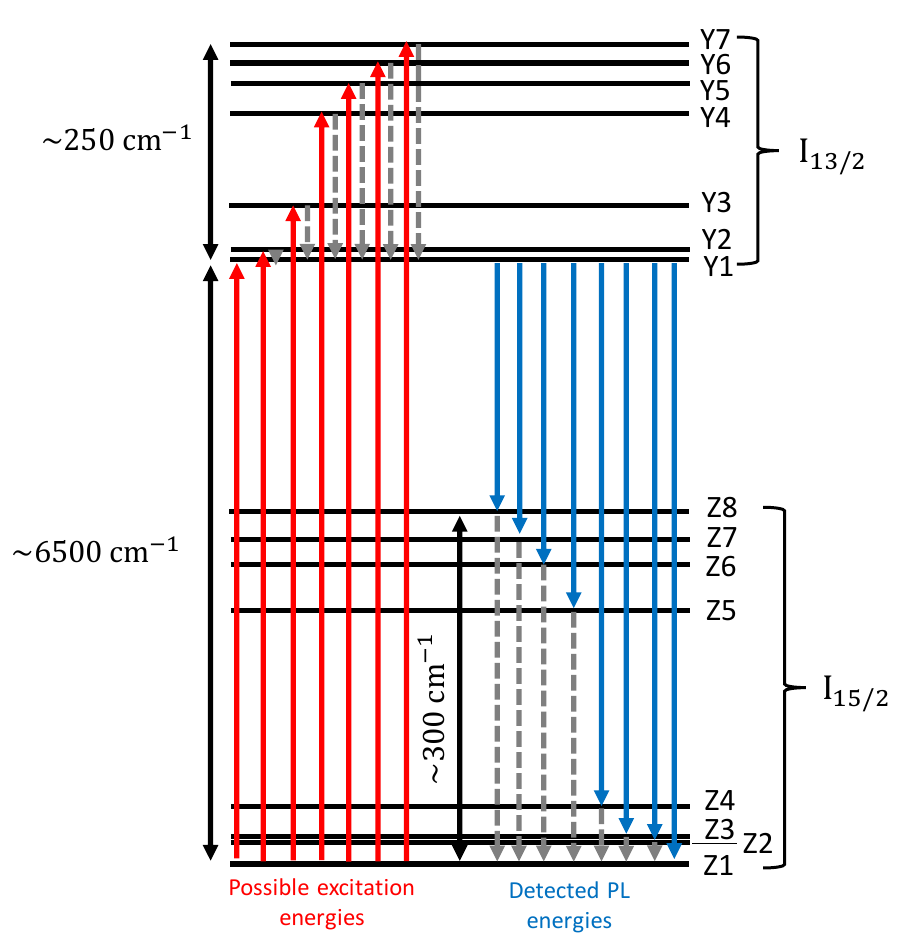}
    \caption{Level scheme with potential optical excitation (red), photon emission (blue), and phonon emission (dashed-grey) paths. Intra-multiplet-energies are to scale, the data is taken from environment 1 (see Table~\ref{T_Levels}).}
    \label{Fig_PLPLE}
\end{figure}
\begin{table*}[t]
\centering
\caption{Summary of assigned $^4\mathrm{I}_{13/2}$ and $^4\mathrm{I}_{15/2}$ energy levels of different identified environments.}
\label{T_Levels}
\begin{tabular}{
    >{\centering\arraybackslash}p{1.5cm} 
    !{\vrule width 1pt} 
    >{\centering\arraybackslash}p{2.875cm} 
    | 
    >{\centering\arraybackslash}p{2.875cm}
    |
    >{\centering\arraybackslash}p{2.875cm}
    |
    >{\centering\arraybackslash}p{2.875cm}
}
\hline
sublevel & \makecell{environment 1 \\ in cm$^{-1}$} & \makecell{environment 2 \\ in cm$^{-1}$} & \makecell{environment 3 \\ in cm$^{-1}$} & \makecell{environment 4 \\ in cm$^{-1}$} \\
\hline
\hline
\rule{0pt}{0.5cm}
Y7 & $\SI{6717.93(6)}{}$ & $\SI{6823.63(6)}{}$ & $\SI{6820.86(6)}{}$ & $\SI{6795.56(44)}{}$ \\
Y6 & $\SI{6701.16(6)}{}$ & $\SI{6738.23(6)}{}$ & $\SI{6749.43(6)}{}$ & $\SI{6753.82(7)}{}$ \\
Y5 & $\SI{6683.38(6)}{}$ & $\SI{6713.40(6)}{}$ & $\SI{6702.75(6)}{}$ & $\SI{6700.05(6)}{}$ \\
Y4 & $\SI{6654.96(6)}{}$ & $\SI{6687.32(6)}{}$ & $\SI{6665.32(6)}{}$ & $\SI{6679.26(6)}{}$ \\
Y3 & $\SI{6573.30(5)}{}$ & $\SI{6614.09(5)}{}$ & $\SI{6603.81(5)}{}$ & $\SI{6652.33(6)}{}$ \\
Y2 & $\SI{6533.033(4)}{}$ & $\SI{6566.81(5)}{}$ & $\SI{6563.70(5)}{}$ & $\SI{6577.77(5)}{}$ \\
Y1 & $\SI{6524.715(4)}{}$ & $\SI{6541.433(4)}{}$ & $\SI{6536.933(4)}{}$ & $\SI{6527.935(4)}{}$ \\
\hline
\rule{0pt}{0.5cm}
Z8 & $\SI{318.3(19)}{}$ & $\SI{437.1(19)}{}$ & $\SI{445.2(19)}{}$ & $\SI{362.5(19)}{}$ \\
Z7 & $\SI{293.8(19)}{}$ & $\SI{336.3(19)}{}$ & $\SI{360.8(19)}{}$ & $\SI{294.8(19)}{}$ \\
Z6 & $\SI{269.5(19)}{}$ & $\SI{315.8(19)}{}$ & $\SI{294.3(19)}{}$ & $\SI{268.0(19)}{}$ \\
Z5 & $\SI{227.7(19)}{}$ & $\SI{262.8(19)}{}$ & $\SI{241.2(19)}{}$ & $\SI{224.1(19)}{}$ \\
Z4 & $\SI{51.62(11)}{}$ & $\SI{104.69(11)}{}$ & $\SI{90.34(11)}{}$ & $\SI{65.32(11)}{}$ \\
Z3 & $\SI{25.84(11)}{}$ & $\SI{72.52(11)}{}$ & $\SI{58.32(11)}{}$ & $\SI{31.92(11)}{}$ \\
Z2 & $\SI{20.13(11)}{}$ & $\SI{39.17(11)}{}$ & $\SI{34.04(11)}{}$ & $\SI{20.01(11)}{}$ \\
Z1 & $\SI{0.00}{}$ & $\SI{0.00}{}$ & $\SI{0.00}{}$ & $\SI{0.00}{}$ \\
\end{tabular}
\end{table*}
\begin{figure*}[t]
    \centering
    \includegraphics[page=1,scale=1.2]{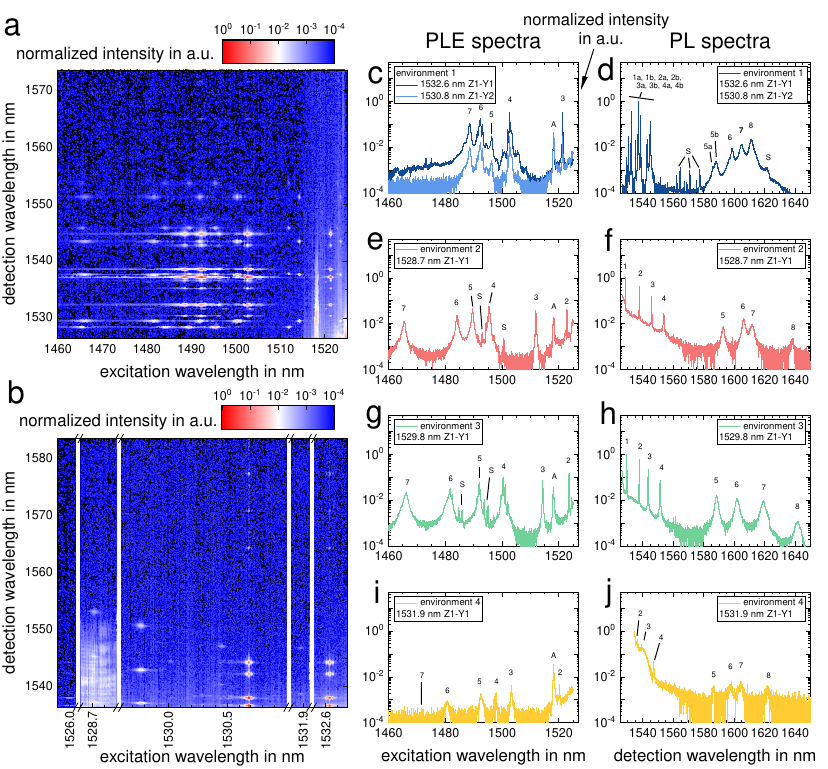}
    \caption{Site-selective \ac{PLE} and \ac{PL} measurements; \textbf{a} and \textbf{b}: Logarithmic-scaled \ac{PLE} maps ; \textbf{c}, \textbf{e}, \textbf{g} and \textbf{i}: Logarithmic \ac{PLE} spectra (cuts from subfigure a) of four assigned environments; \textbf{d}, \textbf{f}, \textbf{h} and \textbf{j}: Logarithmic \ac{PL} spectra of four assigned environments. The numbers labeling the peaks correspond for \ac{PLE} spectra to Y-level assignments and for \ac{PL} spectra to Z-level assignments.The letters S and A are used to indicate that a peak corresponds to a satellite peak and scattered laser light in the spectrometer, respectively.}
    \label{Fig_PLE_Meas}
\end{figure*}
In this section, we present a \ac{PLE} and \ac{PL} study revealing at least four distinct environments experienced by Er$^{3+}$.
For each of the four environments, we identify the energies of all seven $^4\mathrm{I}_{13/2}$ and eight $^4\mathrm{I}_{15/2}$ sublevels, which are split by crystal field. Figure~\ref{Fig_PLPLE} shows a schematic of Erbium $^4\mathrm{I}_{13/2}$ and $^4\mathrm{I}_{15/2}$ multiplets in CaWO$_4$ with potential excitation and relaxation paths. As we measured at a temperature of $\SI{4.3}{K}$, only the Z1 levels should be populated, and spontaneous phonon emission remains as a relaxation mechanism from higher to lower levels within a multiplet \cite{Luo.2020}. These phonon relaxation processes are typically in the order of $\leq \SI{100}{ps}$. As the radiative lifetime from the $^4\mathrm{I}_{13/2}$ multiplet to the $^4\mathrm{I}_{15/2}$ multiplet is on the order of milliseconds, optical relaxation from the $^4\mathrm{I}_{13/2}$ multiplet is expected to predominantly occur from the Y1 sublevel \cite{Hufner.1978}. Details of the experimental setup and sample can be found in Section~\ref{S_Setup}.
Figure~\ref{Fig_PLE_Meas}a and b show the \ac{PLE} intensity in logarithmic scale as a function of the excitation wavelength and detection wavelength, while Figure~\ref{Fig_PLE_Meas}c-j highlight specific \ac{PLE} and \ac{PL} spectra of four distinct  environments. For Figure~\ref{Fig_PLE_Meas}a the spectrometer grating was centered at wavelength of $\SI{1550}{nm}$, while the excitation was swept between $\SI{1460}{nm}$ to $\SI{1525}{nm}$ in steps of $\SI{25}{pm}$. The integration time was $\SI{5}{s}$ with 3 averages and an excitation power of $\SI{0.26}{mW}$. The excitation was left-handed circularly polarized. For Figure~\ref{Fig_PLE_Meas}b the excitation laser was swept from $\SI{1525}{nm}$ to $\SI{1533}{nm}$ in step sizes of $\SI{1}{pm}$, while the detection window was centered around $\SI{1560}{nm}$. For this measurement, we used a cross-polarized excitation-detection scheme to reduce laser scattering. The integration time was $\SI{5}{s}$ with 10 averages and an excitation power of $\SI{0.5}{mW}$. Both maps show a multiplicity of excited emissions with different widths, intensities, and patterns. Compared to the only site known for Er:CaWO$_4$ \cite{Enrique.1971} with charge compensation species, not only satellite peaks from this site but, rather distinct other environments are present.\\
In Figures~\ref{Fig_PLE_Meas}c, e, g and i, we highlight four cuts along the excitation wavelength (\ac{PLE} spectra) of Figure~\ref{Fig_PLE_Meas}a. These four cuts correspond to four different environments, for each of which we could fully assign all Y and Z energy levels. The indicated numbering of the peaks denotes the assigned Y-levels. The summary of the assignment is in Table~\ref{T_Levels}. Environment 1 shows the highest peak intensity and its energy level aligns with the literature values of Er:CaWO$_4$ with charge compensation \cite{Enrique.1971}. Figures~\ref{Fig_PLE_Meas}d, f, h and j show the corresponding \ac{PL} spectra of the different environments. For environment 3, the excitation was to the Y3 level and for environment 4 to the Y1 level. At these excitation wavelengths, only the specific environment is excited. The numbering of the peaks denotes the assigned Z-levels. Although environment 1 is the reported literature environment \cite{Enrique.1971} and shows the highest intensity, it has the most features and satellites within \ac{PLE} and \ac{PL} spectra. Although there is overlap in the excitation energy when exciting environment 1 and the other satellite environments, the actual \ac{PL} in Subfigures d, f, h and j suggest these satellites have a smaller energy displacement relative to environment 1 than is seen by environments 2 - 4. Thus, environment 1 and neighbouring satellites seem to experience rather the assumed long-range charge compensation \cite{Rancic.2022, Kiel.1970}, whereas environments 2 - 4 may experience a rather nearby charge compensation. Furthermore, only Y1 to Z multiplet transitions are visible for environment 2 - 4, as discussed for Figure~\ref{Fig_PLPLE}. However, environment 1 shows additional emissions from Y2 to the Z levels. With a Y1Y2 spacing of approximately $\SI{250}{GHz}$ the Y2 population could still originate from an occupied phonon mode. As this feature was also seen in implanted Er:CaWO$_4$ \cite{Ourari.192023}, we can exclude a growth-related phenomenon.
Finally, the transitions involving higher Y and Z levels are broadened through the spontaneous phonon emission. This is discussed in more detail in Section~\ref{Sec_FWHM}.
\section{Dipole Orientation and Selection Rules}\label{Sec_Pol}
\begin{figure}[ht]
    \centering
    \includegraphics[width=\columnwidth,trim=0 30 0 120,clip]{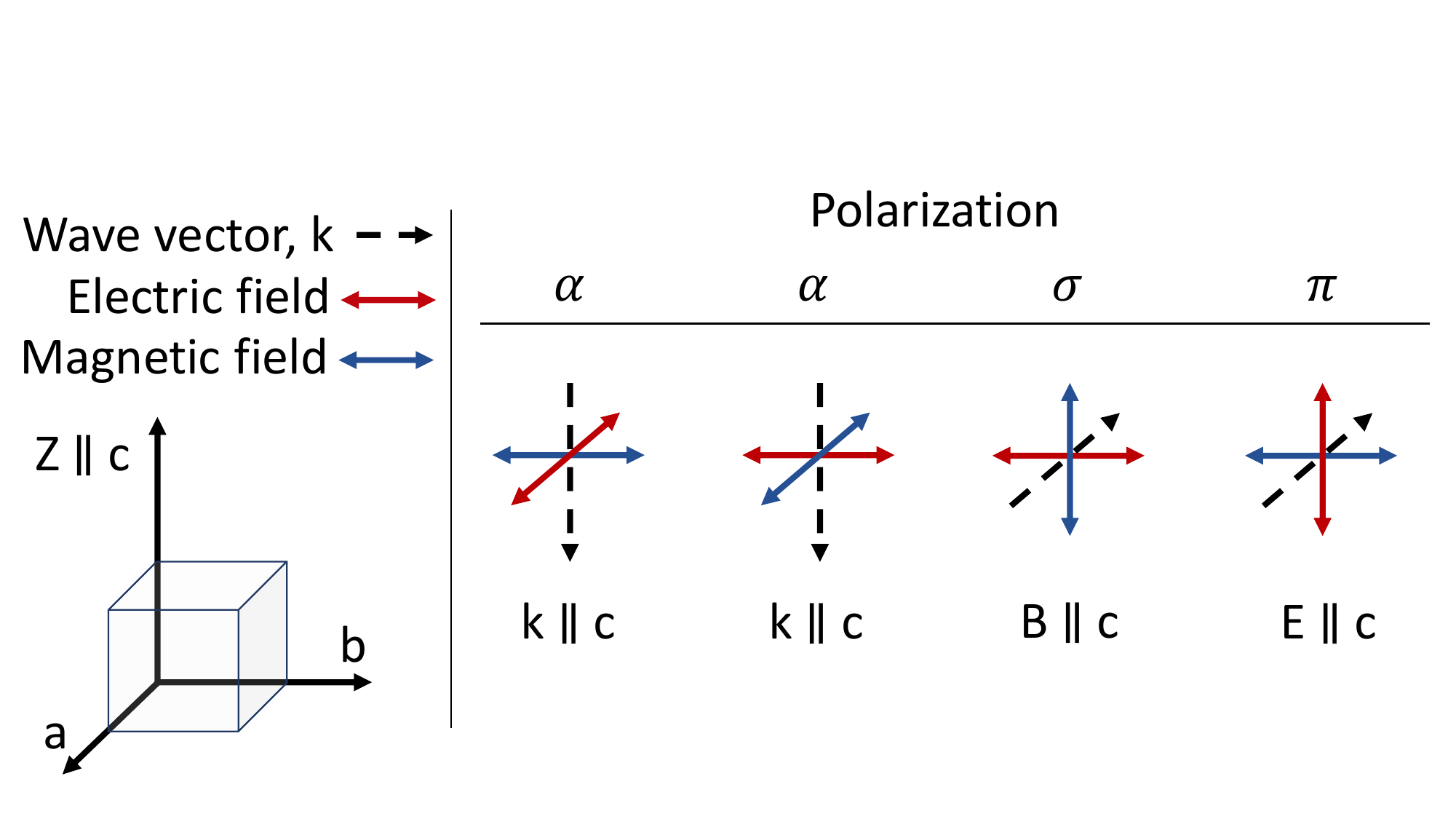}
    \caption{Light polarization relative to the CaWO$_4$ crystal axes with E and B being the electric and magnetic field component.}
    \label{Fig_Pol_Orien}
\end{figure}
\begin{figure*}[ht]
    \centering
    \includegraphics[scale=1.2,trim=0 70 0 0,clip]{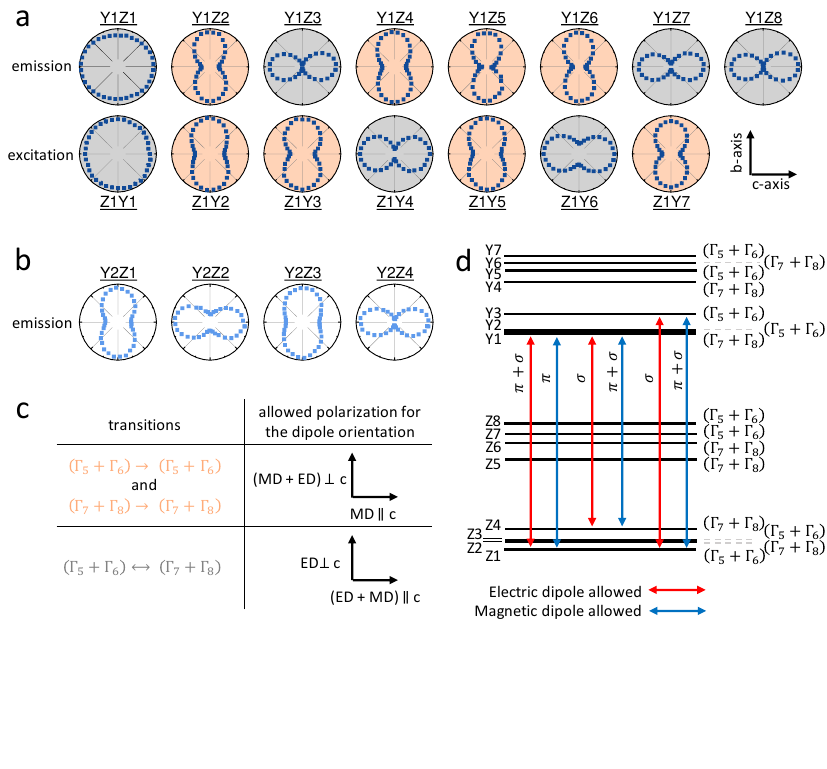}
    \caption{Polarization dependent analysis of environment 1. \textbf{a}: emission and excitation polarization relative to crystal orientation and $^4\mathrm{I}_{15/2}$ and $^4\mathrm{I}_{13/2}$ multiplet energy levels; \textbf{b}: detection polarization of Y2 emissions; \textbf{c}: resulting dipole selection rules from crystal field theory; \textbf{d}: identified irreducible representation of energy levels with indication of the three existing selection rules. The error bars are smaller than the data points.}
    \label{Fig_Pol_En1}
\end{figure*}
\begin{figure*}[t]
    \centering
    \includegraphics[scale=1.2,trim=0 40 0 0,clip]{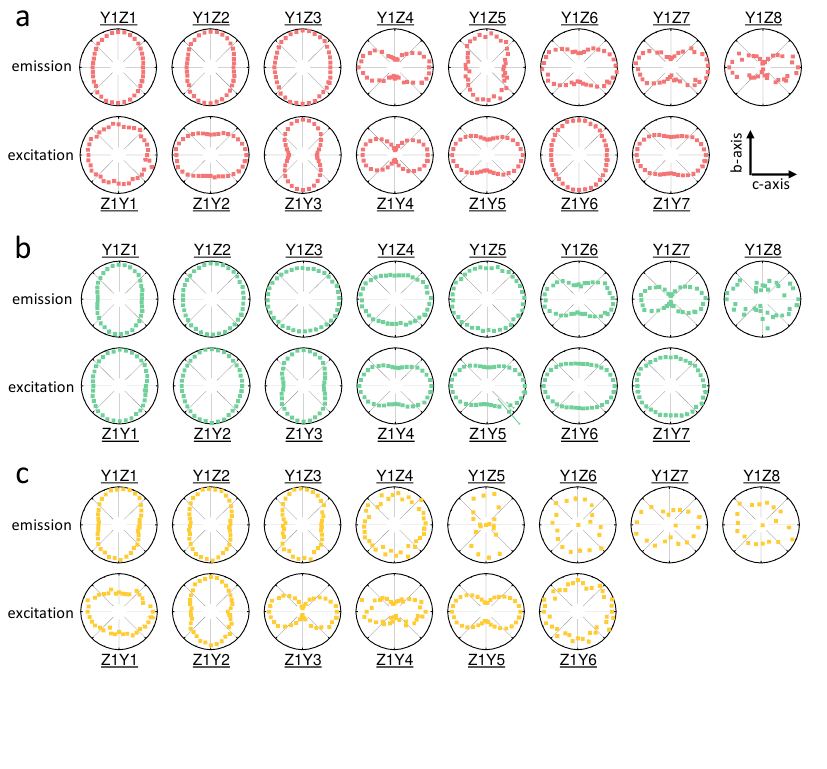}
    \caption{Emission and excitation polarization of environments 2 - 4 with respect to the crystal orientation; \textbf{a}: environment 2; \textbf{b}: environment 3; \textbf{c}: environment 4. The error bars are smaller than the data points.}
    \label{Fig_Pol_En2}
\end{figure*}
In this section we present the measured emission and absorption polarization patterns of $^4\mathrm{I}_{13/2}$ to $^4\mathrm{I}_{15/2}$ transitions and assign based on applied crystal field theory the irreducible representation of involved energy levels. With the assigned states, we can specify if a transition is \ac{ED} or \ac{MD} allowed under a certain polarization.\\
\begin{table}[ht]
\centering
\caption{\ac{ED} selection rules for S$_4$ point group symmetry}
\label{T_ED_Sel}
\begin{tabular}{c||cccc}
  S$_4$ & $\Gamma_5$ & $\Gamma_6$ & $\Gamma_7$ & $\Gamma_8$ \\
\hline
\hline
$\Gamma_5$ & $-$ & $\sigma + \alpha$ & $\pi$ & $\sigma + \alpha$ \\
$\Gamma_6$ & $\sigma + \alpha$ & $-$ & $\sigma + \alpha$ & $\pi$ \\
$\Gamma_7$ & $\pi$ & $\sigma + \alpha$ & $-$ & $\sigma + \alpha$ \\
$\Gamma_8$ & $\sigma + \alpha$ & $\pi$ & $\sigma + \alpha$ & $-$ \\
\end{tabular}
\end{table}
\begin{table}[ht]
\centering
\caption{\ac{MD} selection rules for S$_4$ point group symmetry}
\label{T_MD_Sel}
\begin{tabular}{c||cccc}
  S$_4$ & $\Gamma_5$ & $\Gamma_6$ & $\Gamma_7$ & $\Gamma_8$ \\
\hline
\hline
$\Gamma_5$ & $\sigma$ & $\pi + \alpha$ & $-$ & $\pi + \alpha$ \\
$\Gamma_6$ & $\pi + \alpha$ & $\sigma$ & $\pi + \alpha$ & $-$ \\
$\Gamma_7$ & $-$ & $\pi + \alpha$ & $\sigma$ & $\pi + \alpha$ \\
$\Gamma_8$ & $\pi + \alpha$ & $-$ & $\pi + \alpha$ & $\sigma$ \\
\end{tabular}
\end{table}
As derived in Section~\ref{S_Cal_I1315}, the $^4\mathrm{I}_{13/2}$ crystal field levels have an S$_4$ point group symmetry and can be represented by a combination of 4x ($\Gamma_5+\Gamma_6$) and 3x ($\Gamma_7+\Gamma_8$). And $^4\mathrm{I}_{15/2}$ can be represented by 4x ($\Gamma_5+\Gamma_6$) and 4x ($\Gamma_7+\Gamma_8$).
Additionally, the polarization selection rules can be worked out as described in~\ref{S_Dip_Selec} using Equation~\ref{E_Selection} and are summarized in Table~\ref{T_ED_Sel} and~\ref{T_MD_Sel}. The light polarization orientations with respect to the CaWO$_4$ crystal axis are summarized in Figure~\ref{Fig_Pol_Orien}.
\begin{align}
    \Gamma_i \otimes \Gamma_f^* &= \sum_{j}\Pi_j \label{E_Selection}
\end{align}
Figure~\ref{Fig_Pol_En1} shows measured PL intensity of environment~1 as a function of the emission and absorption polarization in polar coordinates and relates the derived selection rules to the different transitions. Error-bars are in most cases smaller than the data points. In Figure~\ref{Fig_Pol_En1}a, the top row of polarization plots shows the dependence on polarization for emission from Y1 to the lower multiplet and the lower row presents the polarization dependence of the excitation from Z1 to the upper multiplet. The colored background indicates the resulting polarization selection rules discussed in Subfigures~\ref{Fig_Pol_En1}c and d. In the upper row, the main polarization axis for Y1Z2, Y1Z4, Y1Z5, and Y1Z6 is the b-axis, and for Y1Z3, Y1Z7, and Y1Z8 the c-axis. All of these measurements show a bow-tie shape, indicating a reasonable signal suppression in the minor polarized axis. Y1Z1 shows a slight polarization along the c-axis, with less suppression along the b-axis. This 2x4 pattern satisfies the derived 4x($\Gamma_5+\Gamma_6$) and 4x ($\Gamma_7+\Gamma_8$) splitting. In the excitation plots Z1Y2, Z1Y3, Z1Y5 and Z1Y7 are mainly polarized in the b-axis and Z1Y4 and Z1Y6 mainly in the c-axis. Z1Y1 shows a slight polarization along the b-axis. Thus, the excitation polarization does not satisfy the 4x ($\Gamma_5+\Gamma_6$) and 3x ($\Gamma_7+\Gamma_8$) splitting. In order to evaluate this further, we present the detection polarization of the first four transitions from Y2 level in Figure~\ref{Fig_Pol_En1}b. All four show an inverted polarization compared with the detection polarization Y1Z1, Y1Z2, Y1Z3, and Y1Z4 with suppression of the minor axis. This indicates, Y2 as a different irreducible representation as the state Y1. The table in Figure~\ref{Fig_Pol_En1}c reduces the derived \ac{ED} and \ac{MD} selection rules for S$_4$ point group symmetry to a pair of different transitions. When there is a transition between states of the same irreducible representation, only the \ac{ED} transition is allowed perpendicular to the c-axis, whilst the \ac{MD} transition is allowed in all orientations. On the other hand, if there is a transition between states of different irreducible representation, then the \ac{ED} transition is allowed in all orientations but the \ac{MD} transition is only allowed parallel to the c-axis. We assume, that the combination of both dipoles is stronger than a single allowed dipole. Thus, transitions  with an emission polarization axis strongest along the c-axis should correspond to a change of irreducible representations of the initial and final level, whereas, an axis of strongest polrization perpendicular to the c-axis, corresponds to no change in irreducible representation. For a given Y2-Z level transition, the Y1-Z level transition follows the alternative selection rule. This indicates that the irreducible representation of the Y1 level is not the same as for Y2. Hence, Y1Z1 must be of the opposite main polarization of Y2Z1, which is along the c-axis. Z1Y2 and Y2Z1 are equal in orientation perpendicular to the c-axis. This proofs the requirement, that the initial and final state are equal no matter if the transition is directly excited or if we look at the emission of that transition.
This also shows that the dipole of Y1Z1 and Z1Y1 must be similar oriented and opposite to the Y2 level. Thus, this transition needs to be mainly polarized along the c-axis, due to crystal field theory. The reduced suppression along the minor axis could originate from the vacancy-rich environment. The comparison to charge-balanced Er:CaWO$_4$ via co-doping is so far missing and would need further studies. \\
With this, we can assign all transitions with a grey background of Figure~\ref{Fig_Pol_En1}a to transitions between states of different irreducible representations and the remaining transitions to transitions between states with the same irreducible representation. Figure~\ref{Fig_Pol_En1}d shows a level scheme of the studied energy levels with assigned irreducible representation and three examples of the two types of allowed dipole transitions. As the $^4\mathrm{I}_{13/2}$ multiplet consists of 4x ($\Gamma_5+\Gamma_6$) and 3x ($\Gamma_7+\Gamma_8$) Z1Y2, Z1Y3, Z1Y5 and Z1Y7 are of the nature $(\Gamma_5+\Gamma_6) \xleftrightarrow{} (\Gamma_5+\Gamma_6)$, which means all five included energy levels can be represented by ($\Gamma_5+\Gamma_6$). Then Y1, Y4 and Y6 can be represented by ($\Gamma_7+\Gamma_8$). As we now know the nature of Y1, we can assign Z2, Z4, Z5 and Z6 to ($\Gamma_7+\Gamma_8$) and Z3, Z7 and Z8 to ($\Gamma_5+\Gamma_6$).
Figure~\ref{Fig_Pol_En2} shows the emission and excitation polarization dependencies of environment 2 - 4 relative to the crystal orientation. All transitions in each environment show, in general less restricted dipole selection rules. With the exception of transitions Y1Z1, which seems to be better restricted compared to environment 1 (see Figure~\ref{Fig_Pol_En1}a). Assuming environments 2 and 3 possess as well a S$_4$ point group symmetry, we assigned all irreducible representation and summarised them in Table~\ref{T_IrRep_23}. As for environment 3 the suppression of transitions Y1Z2, Y1Z4, Y1Z5, Z1Y2 and Z1Y7 is low, the assignment for these transitions is subject to potential errors. For environments 2 and 3 the irreducible representations are quite different compared to environment 1. However, environments 2 and 3 only differ in the Y2 and Y6 energy levels, which could result from a similar vacancy distortion and could also explain similarities of the \ac{PL} and \ac{PLE} spectra in Figure~\ref{Fig_PLPLE}. For environment 4 in Subfigure c, this is not possible, as the signal-to-noise ratio is too low.
\begin{table}[ht]
\centering
\caption{Summary of irreducible representation of $^4\mathrm{I}_{13/2}$ and $^4\mathrm{I}_{15/2}$ sublevels of environment 2 and 3}
\label{T_IrRep_23}
\begin{tabular}{
    >{\centering\arraybackslash}p{1.5 cm} 
    !{\vrule width 1pt} 
    >{\centering\arraybackslash}p{3.2 cm} 
    | 
    >{\centering\arraybackslash}p{3.2 cm} 
}
\hline
\rule{0pt}{0.2cm}
sublevel & environment 2 & environment 3\\
\hline
\hline
\rule{0pt}{0.5cm}
Y7 & ($\Gamma_5+\Gamma_6$) & ($\Gamma_5+\Gamma_6$) \\
Y6 & ($\Gamma_5+\Gamma_6$) & ($\Gamma_7+\Gamma_8$) \\
Y5 & ($\Gamma_5+\Gamma_6$) & ($\Gamma_5+\Gamma_6$) \\
Y4 & ($\Gamma_5+\Gamma_6$) & ($\Gamma_5+\Gamma_6$) \\
Y3 & ($\Gamma_7+\Gamma_8$) & ($\Gamma_7+\Gamma_8$) \\
Y2 & ($\Gamma_7+\Gamma_8$) & ($\Gamma_5+\Gamma_6$) \\
Y1 & ($\Gamma_7+\Gamma_8$) & ($\Gamma_7+\Gamma_8$) \\
\midrule
Z8 & ($\Gamma_5+\Gamma_6$) & ($\Gamma_5+\Gamma_6$) \\
Z7 & ($\Gamma_5+\Gamma_6$) & ($\Gamma_5+\Gamma_6$) \\
Z6 & ($\Gamma_5+\Gamma_6$) & ($\Gamma_5+\Gamma_6$) \\
Z5 & ($\Gamma_7+\Gamma_8$) & ($\Gamma_7+\Gamma_8$) \\
Z4 & ($\Gamma_5+\Gamma_6$) & ($\Gamma_5+\Gamma_6$) \\
Z3 & ($\Gamma_7+\Gamma_8$) & ($\Gamma_7+\Gamma_8$) \\
Z2 & ($\Gamma_7+\Gamma_8$) & ($\Gamma_7+\Gamma_8$) \\
Z1 & ($\Gamma_7+\Gamma_8$) & ($\Gamma_7+\Gamma_8$) \\
\end{tabular}
\end{table}
\section{Relaxation of Excited States}\label{Sec_FWHM}
\begin{figure*}[t]
    \centering
    \includegraphics[page=1,scale=1.2]{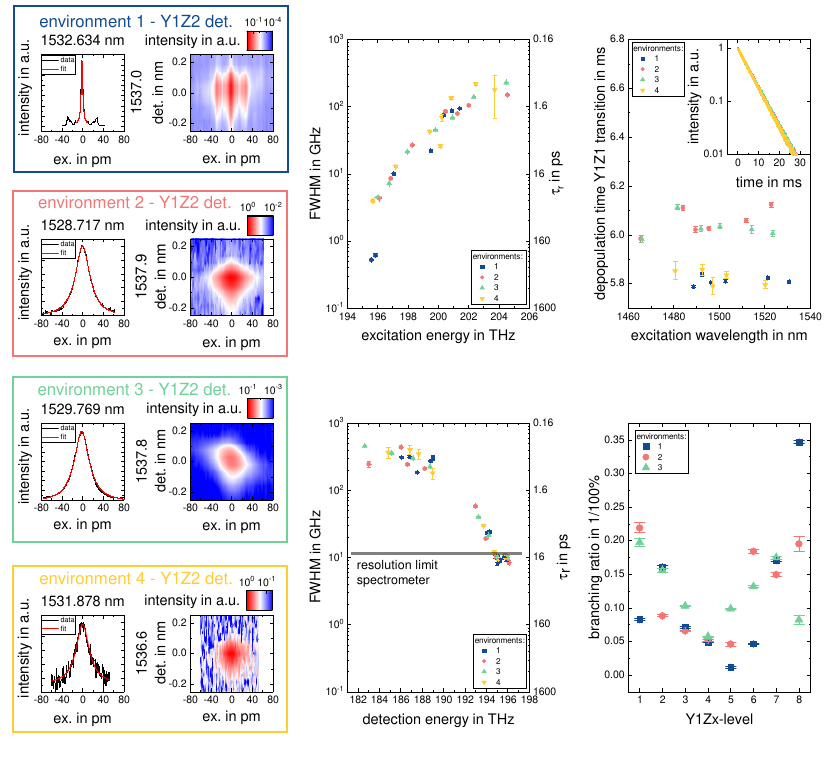}
    \caption{Broadening through different decay paths; \textbf{a}, \textbf{b}, \textbf{c} and \textbf{d}: Absorption spectra of environments Z1Y1 transition line width detection of the Y1Z2 emission (excitation (ex.) and detection (det.)); \textbf{e} and \textbf{f}: Lorentzian fit \ac{FWHM} evolution along higher Y and Z levels. The spectrometer resolution limit for a central wavelength of $\SI{1550}{nm}$ is shown; \textbf{g}: Y1Z1-level depopulation time dependence of different Y level excitation; \textbf{h}: Branching ratio of depopulation paths.}
    \label{Fig_FWHM}
\end{figure*}

In this section, we discuss the line broadening for each transition, compare it to the optical lifetime and discuss the depopulation paths.\\
Figure~\ref{Fig_FWHM}a, b, c and d highlights the absorption spectra of the Z1-Y1 transitions for environments 1 - 4 together with \ac{PLE} maps of the Y1Z2 transitions. For environment 1, the figure shows a central linewidth of $\Delta_{1,\text{inh.}}=\SI{524(9)}{MHz}$ with hyperfine features, while for environment 2 - 4 the peaks are broadened without a similar feature to $\Delta_{2,\text{inh.}}=\SI{4.35(4)}{GHz}$, $\Delta_{3,\text{inh.}}=\SI{4.44(4)}{GHz}$ and $\Delta_{4,\text{inh.}}=\SI{3.94(24)}{GHz}$. The hyperfine feature represents the splitting of the $\SI{22.869}{\%}$ natural abundant $^{167}$Er \cite{Ourari.192023, Rancic.2022, Haynes.op.2014,Bertaina.2007,Mason.1968}. The increased inhomogenous broadening of environments 2 - 4 indicates a larger local disorder through, for instance, strain or defects. In addition, they emit different at different excitation wavelengths. In particular, environment 3's emission wavelengths decrease with increasing excitation wavelength. This can also be related to an effect of higher local disorder \cite{Stevenson.2022}.\\
Figure~\ref{Fig_FWHM}e shows the \ac{FWHM} of Lorentzian fits as a function of the energy of the Z1-Y transitions used for excitation for all environments. With increasing Y level number, the linewidth broadens from below $\SI{10}{GHz}$ to above $\SI{100}{GHz}$. This results from the increasing influence of the spontaneous phonon emission compared to the mainly inhomogeneously broadened Y1Z1 transitions \cite{Luo.2020}. As spontaneous phonon emission is a homogeneous broadening mechanism, we can relate the radiative lifetime to it. As a consequence of the energy-time uncertainty principle homogeneously broadened lines must satisfy $\Delta \omega \gtrsim 1/\tau_r$ \cite{Fox.2011}. The lifetime calculated from this is plotted to the right y-axis and the resulting homogeneous broadening represents radiative lifetimes of upper Y levels. They decrease from smaller than $\SI{16}{ps}$ to faster than $\SI{1.6}{ps}$ and align with the literature \cite{Hufner.1978}. The relation of energy broadening and acoustic phonon emission for low temperatures is \cite{Luo.2020}
\begin{align}
    \Delta E_{\text{sp}}(\mathrm{in \: cm^{-1}})&=\sum_{f<\:i}\frac{1}{c}\frac{\left(\omega_0\right)^3D^2}{2\pi\rho\hbar}\left(\frac{1}{\nu_\text{l}^5}+\frac{2}{\nu_\text{t}^5}\right)|\langle f| V_1|i\rangle|^2 
\end{align}
where $D$ is the deformation potential constant, $\rho$ the mass density, $\nu_\text{l}$ the longitudinal and $\nu_\text{t}$ transverse acoustic phonon velocity. All these parameters are independent of the energy level, leaving a relation of the sum over involved final $f$ and initial $i$ state with its energy gap $\omega_0^3$ and first-order electron-phonon interaction potential term $|\langle f| V_1|i\rangle|^2$. We know $\omega_0$ from Section~\ref{Sec_PLE}, however, further work would be needed to identify the unknown parameters in order to derive a value for the interaction potential. Phonon energies in CaWO$_4$ have been observed to be as high as $\SI{912}{cm^{-1}}$ \cite{Porto.1967}. As the energy difference between the lowest sublevel and the highest sublevel of the $^4\mathrm{I}_{13/2}$ multiplet is approximately $\SI{250}{cm^{-1}}$, a transition between any multiplet sublevel can occur to a lower sublevel with the emission of one or a couple of phonons. The equation predicts a non-linear increasing dependency with the involved number of energy levels, which aligns with our observation.\\
Figure~\ref{Fig_FWHM}f depicts the same relation between linewidth/lifetime and energy for emission into different Z levels. The most narrow lines are close to our spectrometer resolution limit of $\SI{90}{pm}$. Besides these transitions, the linewidth has a similar dependence on the energy which can also be attributed to spontaneous phonon emission.\\
Figure~\ref{Fig_FWHM}g shows the depopulation lifetime of the Y1 state measured using pulsed excitation and time-resolved detection. The measurement is taken by exciting to higher Y-levels to avoid any potential radiation trapping \cite{Xie.2021,Luo.2020,Li.1992,Gritsch.2022,Sumida.1994,Bottger.2006}. Additionally, we used a frequency filter with a resolution of $\SI{0.15}{nm}$ in the detection to spectrally filter only on the specific Y1Z1 transition. This is especially important for excitations of higher Y levels, as more than one environment is excited simultaneously (see Figure~\ref{Fig_PLE_Meas}a). The inset shows example decays of all environments where we fitted a single exponential curve with offset. The resulting depopulation times for the environment 2 and 3 are quite similar for higher Y-level excitation and separate for lower to values of Y2 excitation of $\tau_{2,\text{Y2}}=\SI{6.12(1)}{ms}$ and $\tau_{3,\text{Y2}}=\SI{6.01(1)}{ms}$. The depopulation time of environments 1 and 4 are also close with $\tau_{1,\text{Y2}}=\SI{5.81(1)}{ms}$ and $\tau_{4,\text{Y2}}=\SI{5.80(2)}{ms}$. 
In implanted and annealed Er:CaWO$_4$ a lifetime of $\SI{6.3}{ms}$ was measured \cite{Ourari.192023}. Causes for this discrepancy could be the potential lower temperature of this measurement of $\SI{0.47}{K}$ or the excitation of the Y1 line. For a crystal grown by a Czochralski technique with RF heating and Na$^+$ co-doping of 1:1 with $\SI{0.3}{\%}$ Er$^{3+}$ doping, the lifetime at 10 K was measured to be $\SI{6.62}{ms}$ \cite{Cornacchia.2007}. As no photon trapping was considered, this difference can either stem from the potentially less disordered environment or photon trapping due to the higher doping concentration.\\
In summary, all our studied environments show a depopulation lifetime of approximately $6.0 \pm 0.2$ ms. As interstitial incorporation of Er$^{3+}$ is unlikely (see Section~\ref{Sec_Host}) and would additionally have a different local environment, particularly with a lower coordination number \cite{Vries.1998}, we believe that the similarities in lifetime, polarization and spectra of the different environments indicate substitutional incorporation with local distortions through vacancies.\\
Figure~\ref{Fig_FWHM}h shows the Y1 to Z level branching ratio. For this, \ac{PL} is taken in $\sigma$ and $\pi$ detection polarization while the excitation was constantly aligned to the most absorbing axis of the Z1Y3 lines. Additionally, the setup efficiency was recorded using a white light source and applied to the \ac{PL} data. On the corrected data we fitted Lorentzian curves. Then, we sum the areas of these fits up according to the branching ratio of uniaxial crystals (see Equation~\ref{F_Trans_Prop} and~\ref{F_BranchingRat} in Section~\ref{Sec_BranchingRatio}) and present the difference. \\
Environments 2 and 3 show a U-shaped pattern, with their Y1Z1 transition with a relative intensity of $\SI{21.9(8)}{\%}$ and $\SI{19.7(6)}{\%}$. Environment 1 shows a zig-zag pattern with the Y1Z1 transition of $\SI{8.2(1)}{\%}$ and the highest value of the Y1Z8 transitions of $\SI{34.7(2)}{\%}$. For quantum platforms, mainly the Y1Z1 transition is the transition of interest. Thus, the fraction of light emitted into this transition needs to be improved for applications. This can be done via the Purcell-enhancement of ions in a cavity. Taking the alignment of the cavity and Er$^{3+}$ dipole in the uniaxial CaWO$_4$ into account, the branching ratio changes can be described by Equations~\ref{F_Purc_A} and~\ref{F_Beta_Pur}.
\begin{equation}
\begin{split}
    A_P\left(J_1 \rightarrow J_1'\right)&=F_\text{P}^{(\pi)}\frac{1}{3}A^{(\pi)}\left(J_1 \rightarrow J_1'\right)\\
    &\quad+F_\text{P}^{(a)}\frac{1}{3}A^{(a)}\left(J_1 \rightarrow J_1'\right)\\
    &\quad+F_\text{P}^{(b)}\frac{1}{3}A^{(b)}\left(J_1 \rightarrow J_1'\right) \label{F_Purc_A}
\end{split}
\end{equation}
\begin{align}
   \beta_{11}&=\frac{A_\text{P}\left(J_1 \rightarrow J_1'\right)}{A_\text{P}\left(J_1 \rightarrow J_1'\right)+\sum_{j\:>1}A\left(J_1 \rightarrow J_j'\right)} \label{F_Beta_Pur}
\end{align}
The Purcell enhancement is a function of $F_\text{P}\propto\frac{Q}{V}$ \cite{Fox.2011}. Hence, for single ion interactions where high quality factors and low volumes are feasible, the Y1Z1 transition probability $A_{11}\rightarrow 1$. For instance, if $A_\text{P}\left(J_1 \rightarrow J_1'\right)$ of environments 1 is enhanced by 100 or 1000  its branching ratio would increase to approximately $\SI{90}{\%}$ or $\SI{99}{\%}$. However, this is not given for schemes coupling to an ensemble of ions and could limit the overall efficiency. 
\section{Summary}
To summarize, for not charged-balanced Er:CaWO$_4$, we identified several environments and assigned all energy levels of $^4\mathrm{I}_{15/2}$ and $^4\mathrm{I}_{13/2}$ for four environments, where environment 1 matches the literature values. Further, we derived the polarization selection rules using crystal field theory, identified the irreducible representations of environments 1 - 3 assuming an S$_4$ point group symmetry, and showed similarities between environments 2 and 3. To reveal the nature of environments 2 and 3 point group symmetry, the symmetry-dependent g-tensors of these environments could be measured. Finally, we investigated the linewidths and relaxation paths, where for all environments phonon-emission broadening of the lines was comparable. Environments 2 and 3 were especially similar in terms of depopulation time and branching ratio. Overall, we want to draw attention to the additional environments and give a detailed picture for further studies and applications.
\section{Methods}
\subsection{General setup}\label{S_Setup}
The samples under investigation were Er:CaWO$_4$ crystals from SurfaceNet with a 10 ppm natural abundant Erbium doping, of a size of 3 mm x 4.5 mm x 5.5 mm (a x b x c-axis) and polished on all sides. The doping was added during the hybrid Flow-Zone-Czochralski growth. The measurements were taken at $\SI{4.3}{K}$ in Attodry800XS and Attodry2100 closed cycle cryostats. For excitation, a CTL1500 from Toptica locked with a WS7-IR wavemeter was used. Additionally, the power in the current excitation polarization at the setup was stabilized using a home-build PID control and a V1550PA fiber attenuator from Thorlabs. This stabilization was essential for the polarization-dependent analysis in Sections~\ref{Sec_Pol} and~\ref{Sec_FWHM}. The fluorescent signal was detected by a Shamrock 750 spectrograph from Andor using a $\SI{600}{l/mm}$m ruled, $\SI{1200}{nm}$ blaze grating and a Du491A 1.7 InGaAs detector. An SHB05T shutter from Thorlabs is used in the setup before entering the cryostat for background subtraction, which is applied for any data recorded. A 70:30 beam splitter divided excitation and detection paths.
\subsection{Calculation of the \texorpdfstring{$^4$},I\texorpdfstring{$_{13/2}$}, and \texorpdfstring{$^4$},I\texorpdfstring{$_{15/2}$}, Representation}\label{S_Cal_I1315}
When an ion such as Er$^{3+}$ is placed in a crystal the symmetry of the crystal field Hamiltonian is reduced from R3 to the point group symmetry of the location of the ion in the crystal. This point group symmetry of a site provides information on all the symmetry elements of the site. For the S$_4$ symmetry of Er$^{3+}$ in CaWO$_4$ these symmetry elements are E, $C_2$, $S_4$ and $S_4^{-1}$. The meaning of these elements are described in Table~\ref{T_Sym_Elemets}  \cite{Bethe.1929, Koster.1969}.\\

The electronic states have their own symmetry and this symmetry can be given a representation. It is convenient
to write the representations of the state in terms of the irreducible representations of
the point group symmetry. The irreducible representations are the simplest possible
representations of a group from which all other representations can be expressed and
characterizes the behaviour of the state under the symmetry elements of the point group.
For an ion with total angular momentum J, the irreducible representation of the
operator representing full rotational symmetry, R3, has dimensionality (2J+1). When
placed into a crystal, the previously reducible representation can now be reduced to
the irreducible representation of the subgroup associated with the point group symmetry
of the site \cite{Bethe.1929, Dresselhaus.2008, Wybourne.1965}.

To decompose the reducible representations into the irreducible representations, the
characters of the states in the reducible representation need to be rewritten as the sum
of the characters from the irreducible representations of the point group. The character
of a state with total angular momentum, \textit{J}, in R3 point group symmetry for a rotational
operation of angle $\phi$ around an arbitrary axis is given by \cite{Bethe.1929}:
\begin{align}
\chi\left(\phi\right)&=\frac{\sin\left(\left(2J+1\right)\phi/2\right)}{\sin\left(\phi/2\right)}\label{F_Character}\\
    \lim_{\phi \to 0} \chi\left(\phi\right)&\approx\frac{2\left(J+1\right)\phi/2}{\phi/2}=2\left(J+1\right)
\end{align}
In equation~\ref{F_Character}, \textit{J} is the total angular momentum and $\phi$ is the rotation angle associated
with the group element. For example, a C2 group element corresponds to a $\pi$ rotation,
therefore $\phi=\pi$. For Er$^{3+}$, \textit{J} has a non-integer value, meaning for every $\phi$ an additional
$2\pi$ rotation will result in a different character value, $\chi(\phi+2\pi)=-\chi(\phi)$. To account
for the problem of every character being double-valued, a whole rotation around an
arbitrary axis is redefined to be $4\pi$ rather than $2\pi$. In doing so, a crystal double group
is introduced where every element in the single-valued group now includes an equivalent
element with an additional $2\pi$ rotation applied \cite{Bethe.1929}. Further more, because \textit{J} is a half-integer,
whenever a symmetry element corresponds to a $\pi$ rotation, the character element
will be equal to zero. For symmetry elements containing a reflection, the inversion
symmetry of the state also needs to be considered. A reflection can be decomposed into
an inversion and rotation. Equation~\ref{F_Character} gives the rotation component and the inversion
symmetry is given by the parity of the orbitals (f orbitals have odd parity). To use
Equation~\ref{F_Character} to find the character for a reflection symmetry element, $\phi$ is given a value
of $\pi$ and this results in a character value of zero. For the rotation-reflection element S$_4$,
$\phi$ is given a value of $\pi/2$ and the negative of the resultant character is used to account
for the inversion symmetry component \cite{Luo.2020}. \\ \\
For the S$_4$ point group symmetry of our sample, the relevant characters are given in Table~\ref{T_Sym_Elemets} and~\ref{T_CharacterTable} \cite{Koster.1969}. Using these tables and Equation~\ref{F_Character}, the
irreducible representation for the lowest 2 terms of Er$^{3+}$ can be calculated as a linear combination. $^4\mathrm{I}_{13/2}$ and $^4\mathrm{I}_{15/2}$ have a total angular momentum of \textit{J} = 13/2 and 15/2. For the S$_4$ point
group symmetry, the irreducible representations $\Gamma_5$ and $\Gamma_6$ ($\Gamma_7$ and $\Gamma_8$) are each one dimensional and are the complex conjugates of each other. Due to Kramer’s degeneracy, these irreducible representations will appear as a degenerate pair unless a magnetic field or exchange interaction is present to lift the degeneracy \cite{Wybourne.1965}. Just to remind, the character of the irreducible representation under the identity symmetry element, E, returns the dimensionality of the representation.

\begin{table*}[tb]
\centering
\caption{Symmetry elements of a S$_4$ point group symmetry and their descriptions \cite{Bethe.1929, Koster.1969}.}
\label{T_Sym_Elemets}
\begin{tabular}{c|p{10cm}}
\hline
Symmetry element & Description \\
\hline
\hline
\rule{0pt}{0.5cm}
$E$   & The identity operator. In the context of double groups, this operator is equivalent to a 0 or 4$\pi$ rotation around any given axis. \\
$\overline{E}$ & A rotation by $2\pi$ around any given axis \\
$C_2$ & A rotation by $\pi$ about the optical axis \\
$\overline{C_2}$ & A rotation by $3\pi$ about the optical axis \\
$S_4$ & An improper rotation consisting of a rotation of $\pi/2$ about the optical axis and a reflection through the plane perpendicular to the optical axis \\
$\overline{S_4}$ & An improper rotation consisting of a rotation of $2\pi + \pi/2$ about the optical axis and a reflection through the plane perpendicular to the optical axis \\
$S_4^{-1}$ & An improper rotation consisting of a rotation of $-\pi/2$ about the optical axis and a reflection through the plane perpendicular to the optical axis \\
$\overline{S_4^{-1}}$ & An improper rotation consisting of a rotation of $2\pi - \pi/2$ about the optical axis and a reflection through the plane perpendicular to the optical axis \\
\end{tabular}
\end{table*}

\begin{table*}[tb]
\centering
\caption{Character table for S$_4$ group (adapted from \cite{Luo.2020} and \cite{Koster.1969}).}
\label{T_CharacterTable}
\begin{tabular}{c|cccccccc|l}
\hline
 & $E$ & $\overline{E}$ & $S_4^{-1}$ & $\overline{S_4^{-1}}$ & $C_2$ & $\overline{C_2}$&$S_4$& $\overline{S_4}$ & Bases \\
\hline
\hline
\rule{0pt}{0.5cm}
$\Gamma_1$ & 1 & 1 & 1 & 1 & 1 & 1 &1&1 &$R_z$ \\
$\Gamma_2$ & 1 & 1 & -1 & -1 &1&1& -1 & -1 & $z, xy$ \\
$\Gamma_3$ & 1 & 1 & $i$ & $i$ & -1 & -1 &$-i$&$-i$&$x, y, R_x, R_y$ \\
$\Gamma_4$ & 1 & 1 & $-i$ & $-i$ & -1 & -1 &$i$&$i$& $x, y, R_x, R_y$\\
$\Gamma_5$ & 1 & -1 & $e^{i\pi/4}$ &$-e^{i\pi/4}$ & $i$&$-i$ & $-e^{i3\pi/4}$&$e^{i\pi/4}$ & \\
$\Gamma_6$ & 1 & -1 & $-e^{i3\pi/4}$ & $e^{i3\pi/4}$ & $-i$ & $i$ &$e^{i\pi/4}$&$-e^{i3\pi/4}$& \\
$\Gamma_7$ & 1 & -1 & $-e^{i\pi/4}$ & $e^{i\pi/4}$ & $i$&$-i$ & $e^{i3\pi/4}$&$-e^{i\pi/4}$ & \\
$\Gamma_8$ & 1 & -1 & $e^{i3\pi/4}$ & $-e^{i3\pi/4}$ & $-i$&$i$ & $-e^{i\pi/4}$&$e^{i3\pi/4}$ & \\
\end{tabular}
\end{table*}

\begin{table}[tb]
\centering
\caption{Multiplication table for S$_4$ Group. Taken from \cite{Luo.2020}}
\label{T_MultiplicationS4}
\begin{tabular}{cccccccc|c}
\hline
   $\Gamma_1$ & $\Gamma_2$ & $\Gamma_3$ & $\Gamma_4$ & $\Gamma_5$ & $\Gamma_6$ & $\Gamma_7$ & $\Gamma_8$ & $S_4$ \\
\hline
 $\Gamma_1$ & $\Gamma_2$ & $\Gamma_3$ & $\Gamma_4$ & $\Gamma_5$ & $\Gamma_6$ & $\Gamma_7$ & $\Gamma_8$ & $\Gamma_1$ \\
  & $\Gamma_1$ & $\Gamma_4$ & $\Gamma_3$ & $\Gamma_7$ & $\Gamma_8$ & $\Gamma_5$ &  $\Gamma_6$& $\Gamma_2$ \\
  &  &$\Gamma_2$  & $\Gamma_1$ &$\Gamma_8$  &$\Gamma_5$  &$\Gamma_6$  &$\Gamma_7$  & $\Gamma_3$ \\
  &  &  & $\Gamma_2$&  $\Gamma_6$&$\Gamma_7$  &  $\Gamma_8$&$\Gamma_5$  & $\Gamma_4$ \\
  &  &  &  & $\Gamma_3$ &$\Gamma_1$  &  $\Gamma_4$& $\Gamma_2$ & $\Gamma_5$ \\
 &  &  &  &  & $\Gamma_4$ & $\Gamma_2$ & $\Gamma_3$ & $\Gamma_6$ \\
  &  &  &  &  &  &  $\Gamma_3$&  $\Gamma_1$& $\Gamma_7$ \\
 &  &  &  &  &  &  &  $\Gamma_4$& $\Gamma_8$ \\
\end{tabular}
\end{table}

\begin{table*}[tb]
\centering
\caption{Character Table with Irreducible Representations}
\renewcommand{\arraystretch}{1.5} 
\begin{tabular}{c|cccccccc|c|c}
\hline
$J$ & $E$ & $\overline{E}$ & $S_4$ & $\overline{S}_4$ & $C_2$ & $\overline{C}_2$ & $S_4^{-1}$ & $\overline{S}_4^{-1}$ & Irreducible Representation & Number of Levels \\
\hline
\hline
\rule{0pt}{0.5cm}
$\frac{1}{2}$ & 2 & -2 & $-\sqrt{2}$ & $\sqrt{2}$ & 0 & 0 & $\sqrt{2}$ & $-\sqrt{2}$ & $(\Gamma_7 + \Gamma_8)$ & 1 \\
$\frac{3}{2}$ & 4 & -4 & 0 & 0 & 0 & 0 & 0 & 0 & $(\Gamma_5 + \Gamma_6)+(\Gamma_7 + \Gamma_8)$ & 2 \\
$\frac{5}{2}$ & 6 & -6 & $\sqrt{2}$ & $-\sqrt{2}$ & 0 & 0 & $-\sqrt{2}$ & $\sqrt{2}$ & $2(\Gamma_5 + \Gamma_6)+(\Gamma_7 + \Gamma_8)$ & 3 \\
$\frac{7}{2}$ & 8 & -8 & 0 & 0 & 0 & 0 & 0 & 0 & $2(\Gamma_5 + \Gamma_6)+2(\Gamma_7 + \Gamma_8)$ & 4 \\
$\frac{9}{2}$ & 10 & -10 & $-\sqrt{2}$ & $\sqrt{2}$ & 0 & 0 & $\sqrt{2}$ & $-\sqrt{2}$ & $2(\Gamma_5 + \Gamma_6)+3(\Gamma_7 + \Gamma_8)$ & 5 \\
$\frac{11}{2}$ & 12 & -12 & 0 & 0 & 0 & 0 & 0 & 0 & $3(\Gamma_5 + \Gamma_6)+3(\Gamma_7 + \Gamma_8)$ & 6 \\
$\frac{13}{2}$ & 14 & -14 & $\sqrt{2}$ & $-\sqrt{2}$ & 0 & 0 & $-\sqrt{2}$ & $\sqrt{2}$ & $4(\Gamma_5 + \Gamma_6)+3(\Gamma_7 + \Gamma_8)$ & 7 \\
$\frac{15}{2}$ & 16 & -16 & 0 & 0 & 0 & 0 & 0 & 0 & $4(\Gamma_5 + \Gamma_6)+4(\Gamma_7 + \Gamma_8)$ & 8 \\
\end{tabular}
\end{table*}

\subsection{Derivation of Dipole Selection Rules}\label{S_Dip_Selec}
To determine the symmetry specific selection rules, the irreducible representation of the
\ac{ED} and \ac{MD} operators also needs to be reduced from R3 to the
new lower site symmetry irreducible representation. Electric quadrupole transition selection
rules are not considered as calculations have shown these transitions have a rate
on the order of $\SI{0.01}{s^{-1}}$. This is approximately three orders of magnitude smaller than
the calculated magnetic and electric dipole rate of the $^4\mathrm{I}_{13/2}$ to $^4\mathrm{I}_{15/2}$ transition \cite{Dodson.2012}. A
dipole transition can only occur if the product of the initial state’s irreducible representation
and complex conjugate of the final state’s irreducible representation includes the
irreducible representation of the dipole moment operator \cite{Luo.2020}.
\begin{align}
    \Gamma_i \otimes \Gamma_f^* &= \sum_{j}\Pi_j \label{E_Selection2}
\end{align}
In Equation~\ref{E_Selection2}, $\Gamma_i$ and $\Gamma_f$ are the irreducible representation of the initial state and final
state respectively, and $\sum_j \Pi_j$ is the irreducible representation of the dipole operator. The
irreducible representation of the dipole operator can be broken down into the irreducible representations which correspond to either the z or x-y component. This introduces
polarization dependent selection rules between the irreducible representations of the
final and initial states in the crystal field splittings. As the electric dipole operator is a
polar vector with odd parity and the magnetic dipole operator is an axial vector with
even parity, the two dipole operators have different irreducible representations. The irreducible representations of the z and x-y components are summarised in Table~\ref{T_ED_MD_IrRep} \cite{Luo.2020}. Figure~\ref{Fig_Pol_Orien} provides a summary and definition of the three types of polarizations which are used. These definitions are required to define which polarization dependent transitions are allowed or forbidden. These polarizations, $\alpha$, $\sigma$, $\pi$ are all defined relative to the crystal axes. These polarizations are used in Table~\ref{T_ED_Sel} and~\ref{T_MD_Sel}, where $\sigma$ polarization refers to $E\perp c$ and $B\parallel c$, $\pi$ polarization refers to $E\parallel k \parallel$ c and $B\perp c$, and $\alpha$ polarization refers to $E\perp c$ and $B\perp c$, where $\hat{Z}$ is defined along the c crystal axis (the optical axes in the uni-axial crystal CaWO$_4$). \\

Table~\ref{T_MultiplicationS4} is the multiplication table for the irreducible representations for the S$_4$ group. The allowed polarizations for electric and magnetic dipole transitions between two states with an irreducible representation ($\Gamma_7+\Gamma_8$) can be calculated by comparing the result of $(\Gamma_7+\Gamma_8)\otimes(\Gamma_7+\Gamma_8)^{*}=\Gamma_7\otimes\Gamma_7^{*}+\Gamma_8\otimes\Gamma_7^{*}+\Gamma_8\otimes\Gamma_8^{*}=(\Gamma_3+\Gamma_1+\Gamma_4)$ to the irreducible representations of the dipole operators
(found in Table~\ref{T_ED_MD_IrRep}). The final result of $(\Gamma_7+\Gamma_8)\otimes(\Gamma_7+\Gamma_8)^*$ contains the irreducible representation
of: (i) the component of the electric dipole operator in the plane perpendicular to the z-direction (c-axis of the crystal); (ii) the component of the magnetic dipole operator parallel to the z-direction; and (iii) the component of the magnetic dipole operator in the plane perpendicular to the z-direction. Transitions where the electric or magnetic dipole operators match these orientations are allowed transitions. We then compare in this example the orientations of the dipole operators required for a transition to be allowed to the polarizations defined in Figure~\ref{Fig_Pol_Orien}. It is shown that an electric dipole transition is allowed when there is $\alpha$ or $\sigma$ polarization (when the electric field is in the x-y plane, perpendicular
to the crystal c-axis). Whilst, for a magnetic dipole transition, all polarizations are allowed. The electric dipole selection rules for the relevant irreducible representations for S$_4$ site symmetry are given in Table~\ref{T_ED_Sel} and the magnetic dipole selection rules are given in Table~\ref{T_MD_Sel} \cite{Luo.2020}.

\begin{table*}[tb]
\centering
\caption{Selection rules for the S$_4$ point group symmetry.}
\label{T_ED_MD_IrRep}
\begin{tabular}{
    >{\centering\arraybackslash}p{2.5 cm} 
    !{\vrule width 1pt} 
    >{\centering\arraybackslash}p{2.5 cm} 
    | 
    >{\centering\arraybackslash}p{2.5 cm} 
    | 
    >{\centering\arraybackslash}p{2.7 cm} 
    | 
    >{\centering\arraybackslash}p{2.7 cm} 
}
\hline
Point group symmetry & Electric dipole ($z$) & Electric dipole ($x,y$) & Magnetic dipole ($R_z$) & Magnetic dipole ($R_x, R_y$) \\
\hline
\hline
\rule{0pt}{0.5cm}
S$_4$ & $\Gamma_2$ & $\Gamma_3 + \Gamma_4$ & $\Gamma_1$ & $\Gamma_3 + \Gamma_4$ \\
\end{tabular}
\end{table*}

\subsection{Branching Ratio}\label{Sec_BranchingRatio}
For uniaxial crystals as CaWO$_4$ the transition probabilities between multiplets can be calculated as in Equation~\ref{F_Trans_Prop} \cite{Luo.2020}.
\begin{align}
    A\left(J_i \rightarrow J_j\right)&=\frac{1}{3}A^{(\pi)}\left(J_i \rightarrow J_j'\right)+\frac{2}{3}A^{(\sigma)}\left(J_i \rightarrow J_j'\right) \label{F_Trans_Prop}
\end{align}
The radiative lifetime $\tau_r$ and the fluorescent branching ratio $\beta_{ij}$ can be expressed as
\begin{align}
    \tau_r\left(j_i\right)&=\frac{1}{\sum_{J_j}A\left(J_i \rightarrow J_j'\right)}\\
    \beta_{ij}&=\frac{A\left(J_i \rightarrow J_j'\right)}{\sum_{J_j'}A\left(J_i \rightarrow J_j'\right)} \label{F_Beta_Raw}
\end{align}
Moreover, in the derivation of the Fuchtbauer-Ladenburg equation the relation between fluorescent spectra and Einstein coefficient is noted \cite{Luo.2020}
\begin{align}
    \frac{1}{hc}\int_{\text{spectrum}_{2\rightarrow 1}} \lambda I^{q}(\lambda)d\lambda&=GN_uA_{21}^{q} 
\end{align}
where $I^{q}$ denotes the detected fluorescence intensity in polarization q, G the efficiency of the fluorescence detective system and $N_u$ the ion number of the high multiplet. We can now rewrite this, using the center of gravity
\begin{align}
    \frac{1}{hc}\cdot\overline{\lambda} \int_{\text{spectrum}_{2\rightarrow 1}} I^{q}(\lambda)d\lambda&=GN_uA_{21}^{q}
\end{align}
Now the integral is the area of the detected peak in polarization q and $\overline{\lambda}$ the average wavelength of this peak. The detection efficiency is already applied to our spectra, leaving the ion number as the only unknown for calculating $A_{21}^{q}$. However, as we are in a linear region, where a change in power linearly increases or decreases the intensity, $N_u$ is only a proportional factor. Moreover, for environments 2 and 3 only emission of Y1Z1 is detected and for environment 1 the main detection emission is also Y1Z1. Thus, $N_u$ is assumed to be equal for all detected lines. Hence, $N_u$ would cancel out in the Equation~\ref{F_Beta_Raw}. As we took all included spectra with the same power with power stabilization, we can argue that
\begin{align}
    \frac{1}{hc}\cdot\overline{\lambda} \int_{\text{spectrum}_{2\rightarrow 1}} \frac{I^{q}}{G} \left(\lambda\right)d\lambda&\propto A_{21}^{q}
\end{align}
Inserting this into Equation~\ref{F_Beta_Raw} and considering Equation~\ref{F_Trans_Prop} gives us
\begin{widetext}
\begin{align}
    \beta_{1j}&=\frac{\left[\overline{\lambda} \left(\frac{2}{3}\int_{\text{spectrum}_{2\rightarrow 1}} \frac{I^{\sigma}}{G} \left(\lambda\right)d\lambda+\frac{1}{3}\int_{\text{spectrum}_{2\rightarrow 1}} \frac{I^{\pi}}{G} \left(\lambda\right)d\lambda\right)\right]\left(J_1 \rightarrow J_j'\right)}{\sum_{J_j'}\left[\overline{\lambda} \left(\frac{2}{3}\int_{\text{spectrum}_{2\rightarrow 1}} \frac{I^{\sigma}}{G} \left(\lambda\right)d\lambda+\frac{1}{3}\int_{\text{spectrum}_{2\rightarrow 1}} \frac{I^{\pi}}{G} \left(\lambda\right)d\lambda\right)\right]\left(J_1 \rightarrow J_j'\right)} \\
    \beta_{1j}&=\frac{\left[\overline{\lambda}\left(\frac{2}{3}Ar_{\text{cor}}^{\sigma}+\frac{1}{3}Ar_{\text{cor}}^{\pi}\right)\right]\left(J_1 \rightarrow J_j'\right)}{\sum_{J_j'}\left[\overline{\lambda}\left(\frac{2}{3}Ar_{\text{cor}}^{\sigma}+\frac{1}{3}Ar_{\text{cor}}^{\pi}\right)\right]\left(J_1 \rightarrow J_j'\right)} \label{F_BranchingRat}
\end{align}
\end{widetext}
where $Ar_{\text{cor}}^{\sigma}$ is the area of $\sigma$ polarization of the efficiency corrected spectra.
\section{Acknowledgements}\label{S_Acknowledge}
We gratefully acknowledge support from the German Federal Ministry of Education and Research (BMBF) via the funding program Photonics Research Germany (project MOQUA, contract number 13N14846) and project 6G-life, the Bavarian State Ministry for Science and Arts (StMWK) via projects EQAP and NEQUS, the Bavarian Ministry of Economic Affairs (StMWi) via project 6GQT, as well as from the German Research Foundation (DFG) under Germany’s Excellence Strategy EXC-2111 (390814868) and projects PQET (INST 95/1654-1) and MQCL (INST 95/1720-1).
% \bibliographystyle{apsrev4-1}
% \bibliography{bibliography}

\providecommand{\noopsort}[1]{}\providecommand{\singleletter}[1]{#1}%

\end{document}